# Understanding Ion Distribution and Diffusion in Solid Polymer Electrolytes

Ganesh K Rajahmundry and Tarak K Patra[*]
Department of Chemical Engineering
Indian Institute of Technology Madras, Chennai 600036, India

## Abstract

Solid polymer electrolytes (SPEs) - polymer melts with added salts – exhibit ion conduction and high mechanical properties, thus are promising materials for future energy storage devices. The ion conductivity in an SPE is intricately connected to salt ion distribution in the polymer matrix. The relationship between ion diffusion and ion distribution remains unresolved. Here, we conduct coarse-grained molecular dynamics simulations and establish correlations between ion distribution and transport for a phenomenological SPE model. We propose phase diagrams of SPEs as a function of ion pair size, ion concentration and the Bjerrum length of the material. A crossover from a discrete ion aggregate to a percolated ion aggregate is demonstrated as a function of ion pair size in the SPE. Further, we determine ion diffusions for all the phases of the SPE. The work provides important design strategies for controlling ion distribution and enhancing ion conductivity in a polymer matrix.

# Graphical Abstract

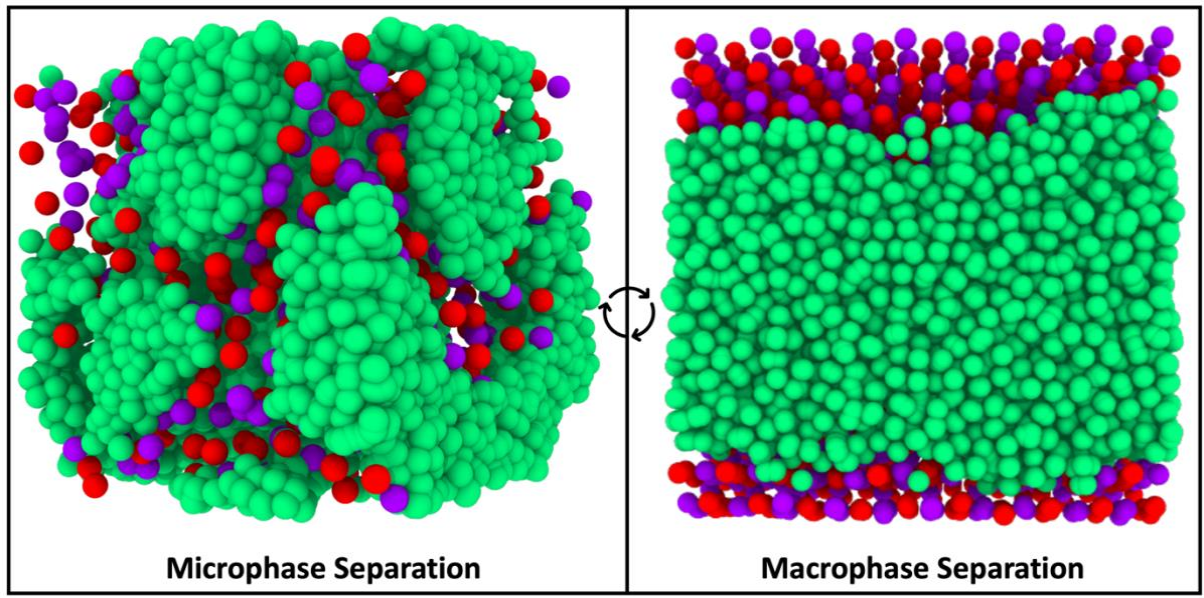

# I. Introduction

Solid polymer electrolytes (SPEs) such as salt-dopped polyethylene oxide (PEO), poly acrylic acid (PAA), poly methyl methacrylate (PMMA), poly acrylonitrile (PAN) are potential candidates to enhance the electrochemical and mechanical stability of batteries.[1–10] Salts within SPEs dissociate into ions that subsequently traverse through the polymer matrix, facilitating ionic conductivity. However, SPEs have a major limitation regarding low ion conduction at room temperature over the currently used liquid electrolytes in batteries. A combination of high ion conductivity and high mechanical properties is desirable for the commercial realization of SPEs for batteries and other electrochemical devices. This remains a substantial challenge in developing SPE for energy storage applications. Recent studies indicate that the ion conductivity can be improved by suppressing the glass transition temperature and/or decoupling ion dynamics from the polymer segmental dynamics of an SPE.[11–14] Ion transport is also found to be intricately correlated to the molecular architecture and resulting microstructure in polymers.[15–20] Often, semicrystalline polymers show better ion-polymer dynamics decoupling and faster ion mobility.[2,21] A comprehensive understanding of microstructure, decoupling of ion motion from polymer dynamics and ion conductivity is lacking.

Molecular simulations can play an important role in understanding above structure-property correlations in an SPE.[22] Towards this end, generic coarse-grained molecular simulations (CGMD) have been progressively used to understand the microstructure and ion transport mechanism in an SPE.[23,24] The CGMD simulations utilize generic bead−spring models of polymers and ions without any chemical specificity. They have played an important role in uncovering the molecular-level structure and dynamics that govern the relationships between microstructure and macroscopic properties in diverse polymeric systems. These CG modeling strategies lack atomistic precision and specific quantitative alignment with chemistry-specific systems. They typically incorporates adjustable parameters to accommodate various chemical compositions and interactions. They enable simulations over extended length and time scales owing to computational effectiveness and offer the advantage of yielding physical insights that are broadly applicable across the entire class of related materials. These methods are considered to be extremely powerful to study the phase behavior and ion transport without any empirical force field and over a wide range of systems including those that have not yet been experimentally realized. They are very useful in understanding the generic relationships between the basic synthetically controllable physical features of polymers/ ions, ion conductivity, and thermomechanical stability. Here we adopt such a CGMD approach to study the ion distribution and dynamics in a polymer melt.

We systematically vary the ion pair size and concentration along with the Bjerrum length of a model SPE, and study the ion distribution and diffusion in the polymer matrix. The Bjerrum length plays an important role in determining the phase behaviour and, thus ion transportation in an SPE. It is commonly used to determine the behavior of charged particles, such as ions, in solutions. The Bjerrum length represents the distance at which the electrostatic energy between two elementary charges in a material is comparable to the thermal energy of the system. When the distance between charges is much larger than the Bjerrum length of the system, the electrostatic interactions between them are effectively screened by the solvent, and the system behaves more like an ideal solution. On the other hand, when the distance between charges is comparable or smaller than the Bjerrum length, the electrostatic interactions dominate and can significantly affect the properties of the system, leading to phenomena such as ion pairing. This has important implication on the phase behavior of an SPE. Here, we report phase diagrams of an SPEs as a function of ion size ratio and the Bjerrum length for different ion concentrations. For low ion concentration, the system exhibits three distinct phases, viz., discrete ion aggregate, percolated ion aggregate and macrophase separation. These percolations appear to strongly impact the ion transport. The anion, which is of the same size as that of the monomer of the polymer, moves faster in a percolating aggregate in comparison to the discrete aggregate. Interestingly, the cation, which is larger in size, move slowly in the percolating aggregate. The cation dynamics is faster in polymer matrix with discrete ion aggregates. For a moderately high concentration, the percolated ion aggregate disappears and the system poses discrete ion aggregate in addition to the macrophase separation. Overall, the present work demonstrates the complex correlation between ion distribution and transport in a polymer melt and show how they are connected to the composition of an SPE.

## II.     Model and Methodology

We use a generic coarse-grained bead-spring side-chain polymer and ion. A schematic representation of the model system is shown in Figure 1a. We consider fully flexible Kremer-Grest (KG) polymer model wherein any two coarse-grained (CG) monomers interact via the Lennard-Jones (LJ ) potential of the form $V(r) = 4\epsilon \left[ \left(\frac{\sigma}{r}\right)^{12} - \left(\frac{\sigma}{r}\right)^{6} \right]$. Here, the size of a monomer is $\sigma$, and the cohesive interaction between two monomers is $\epsilon$. The LJ interaction is truncated and shifted to zero at $r = 2.5\sigma$. In addition, two adjacent monomers in the backbone and side-chains are connected via the finitely extensible non-linear elastic (FENE) potential of the form $E_{FENE}(r_{ij}) = -0.5kr_0^2 \ln\left[1 - \left(\frac{r_{ij}}{r_o}\right)^2\right]$. Here $r_0 = 1.5\sigma$ and $k = 30\epsilon$. The ions are also modeled as LJ particles. In addition to the LJ interaction, the electrostatic interaction between a pair of ion is represented via the Coulomb potential of the form $E_{coul}(r) = \frac{q_i q_j}{4\pi\varepsilon_0 \varepsilon_r r}$. Here, the $\varepsilon_0$ is the permittivity of the vacuum, and the $\varepsilon_r$ is the relative dielectric constant of the medium. The $q_i$ and $q_j$ are the charges of ion $i$ and $j$, respectively. The Bjerrum length of the material is defined as $l_b = q_i q_j / 4\pi\varepsilon\varepsilon_0 k_B T$. The Bjerrum length is varied from 0.5 to 50 in this study. We use the particle-particle particle-mesh (PPPM) method to calculate the long-range electrostatic interaction. The ion-monomer interaction is also modeled via the LJ potential. The size ratio between cation and anions are varied in this study. We consider 225 polymer chains, each consist of 40 coarse-grained particles. We vary the ion concentration, which is defined as the ratio of cations to the polymer beads, from 0.025 to 0.10. The size of the cation is varied from $d_+ = 0.5\sigma$ to $3\sigma$, while the size of the anion remains fixed as $d_- = 1\sigma$. The size ratio of the ions ($D_s = d_+/d_-$) is, therefore, varied from 0.5 to 3.0. The Lorentz Berthelot mixing rule $\sigma_{ij} = (\sigma_i + \sigma_j)/2$ is used for computing the LJ interaction between two particles of different size. We perform simulations for two different ion concentration. Simulation are performed in an isothermal - isobaric ensemble (NPT) with pressure $P = \epsilon/\sigma^3$ and temperature $T = 2\epsilon/k_B$ for $10^7$ steps of equilibration followed by a production run of $10^7$ steps with an integration time step of $0.005\tau$. The temperature and pressure are controlled using the Nosé–Hoover thermostat and barostat, respectively. All the simulations are conducted using the LAMMPS MD package.[25]

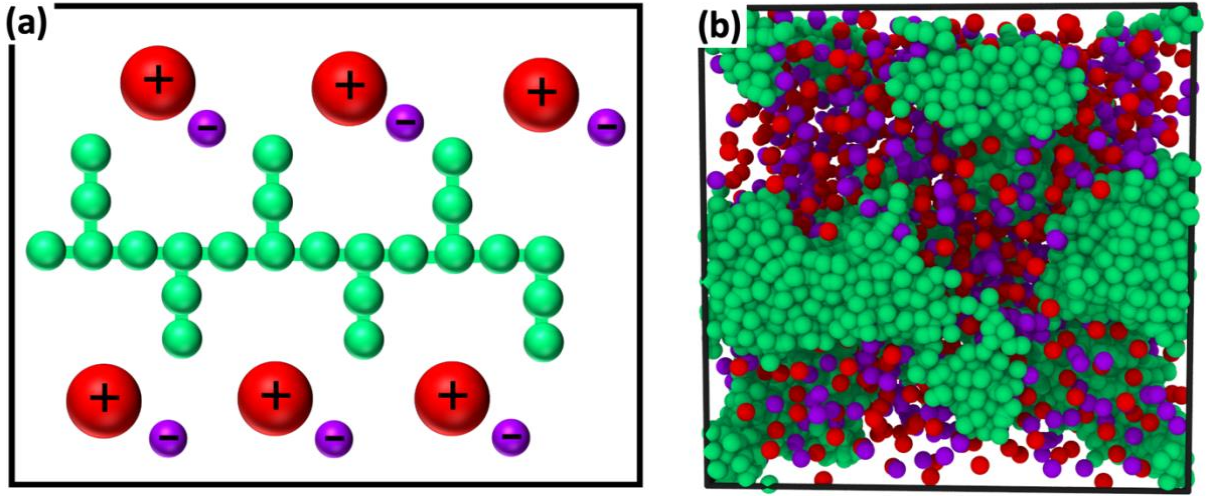

*Figure 1: CGMD of a SPE. A schematic representation of the model system is shown in (a). The size of an anion is fixed and the size of a cation is varied in this study. The polymer has 20 beads in its backbone. Every alternate bead in the backbone is connected with a side-chain of two beads. An equilibrium MD snapshot is shown in (b). The red and purple color correspond to ions, and monomers of polymers are shown as green beads.*

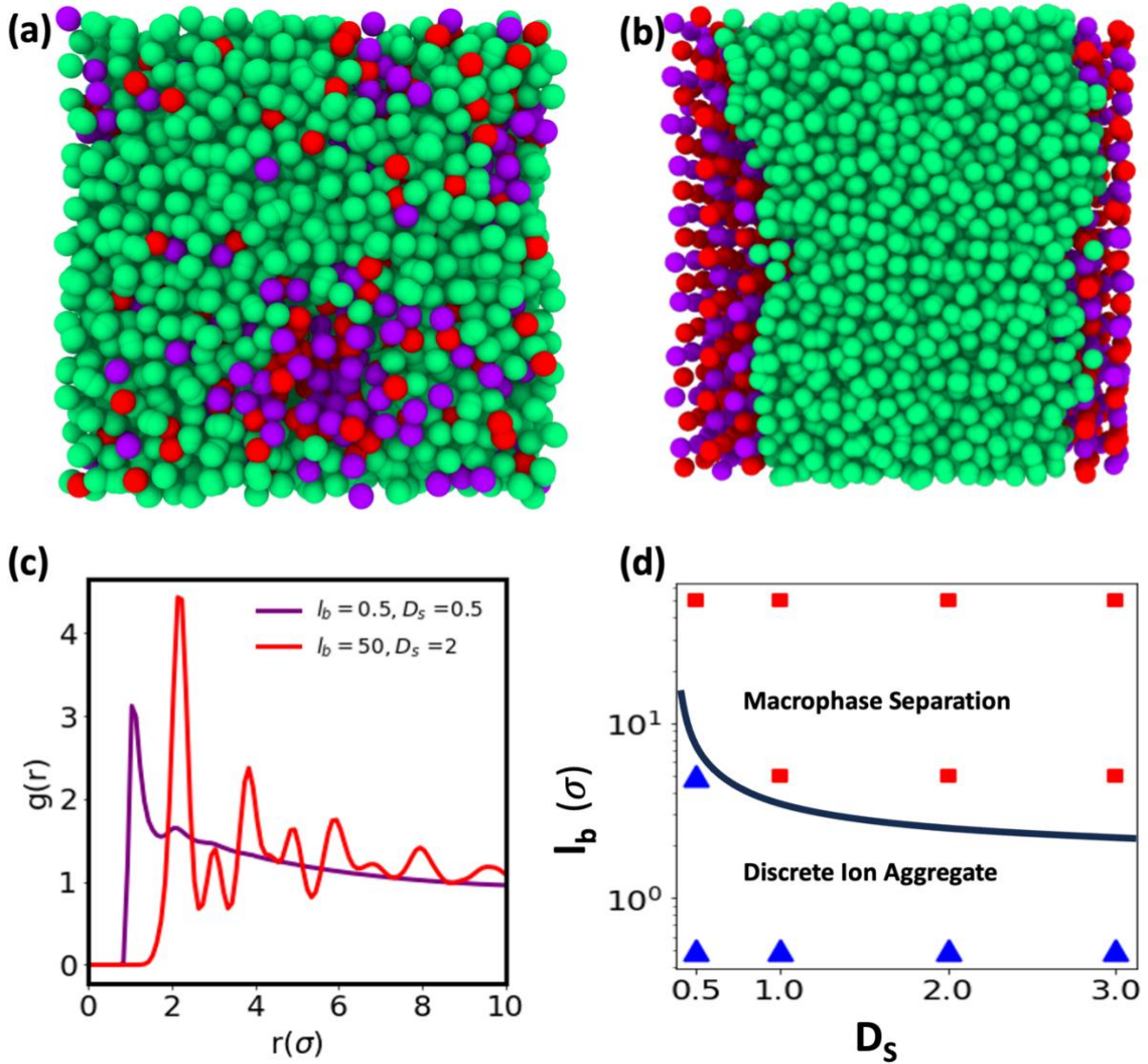

*Figure 2: Phase behavior of SPE for high ion concentration. MD snapshots for $l_b$=0.5 and $D_s$=0.5 are shown in (a). The MD snapshot for $l_b$=50 and $D_s$=2.0 are shown in (b). The green, red and purple colours in (a) and (b) correspond to polymer, cation and anion, respectively. The anion-anion pair correlation functions for the two cases are compared in (c). A phase diagram of all the pashes seen in MD simulations are constructed in (d). The solid line in (d) is a guide to eyes, and not the thermodynamic boundary between the two distinct phases.*

## III. Results and Discussion

We analyze the phase behavior of the SPE in Figure 2 for ion concentration $\rho = 0.10$. For this concentration, there are two distinct phases viz., discrete ion aggregates and macroscopic phase separation. Representative MD snapshots of the two phases are shown in Figure 2a and b, respectively. The corresponding cation-cation pair correlation functions are shown in Figure 2c. It suggests that ion pair correlation is stronger when they are phase separated. The multiple peaks in their *g(r)*, indicates ordered aggregates of ions in this phase. However, the discrete aggregates are amorphous in nature. We construct a phase diagram based on MD simulations, which are conducted for $\rho = 0.10$, and is reported in Figure 2c. For the lower Bjerrum length limit, the systems shows discrete ion aggregates irrespective of the ion pair size. In this limit, thermal energy dominates and leads to mixing of polymers and ion pairs. In this phase, we observe a distribution of ion clusters in the polymer matrix. The cluster size varies from 1 to 51. We note that in recent studies,[26,27] wherein ion-polymer interaction is modelled by a $1/r^4$ potential, similar ion aggregates in SPE are also reported. Although we model ion-polymer interaction using the LJ potential, we essentially capture the phase behaviour which are consistent with these previous studies. As the Bjerrum length increases, the electrostatic interaction among the ion pairs dominates; as a result ions are macroscopically phase separated from the polymer domain. This is evident from the MD snapshot in Figure 2b.

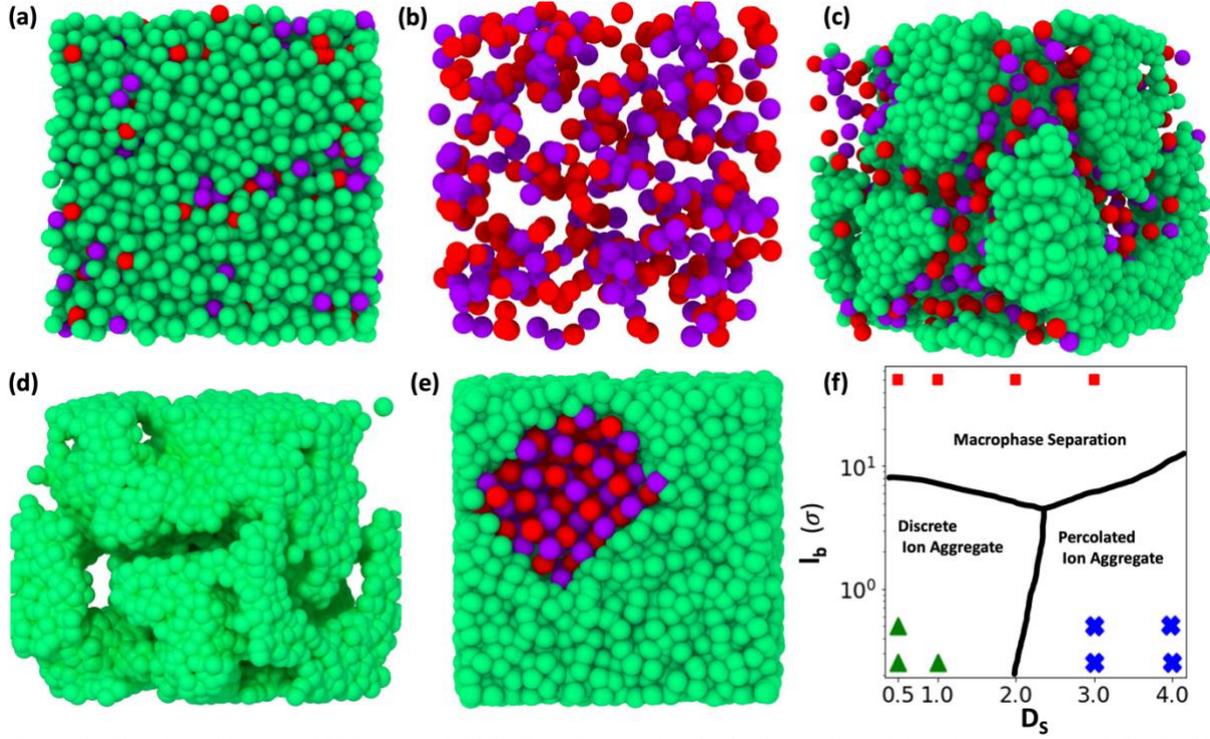

*Figure 3: The phase bhavior of SPE at $\rho = 0.025$. The MD snapshot for $l_b$=0.5 and $D_S$=0.5 is shown in (a) and (b). In (b), polymer beads are deleted for visual inspection of discrete aggregates. The MD snapshots for $l_b$=0.5 and $D_S$=3.0 are shown in (c) and (d). The ions are deleted in (c) for better visual inspection of percolating structure in a polymer matrix. The MD snapshot of macrophase separation at $l_b$=50 and $D_S$=2 is shown in (e). The red and purple color represent cation and anion respectively. Monomers of polymers are shown as green beads. The phase diagram as a function of $l_b$ and $D_s$ is reported in (f). The solid line (f) is a guide to eyes.*

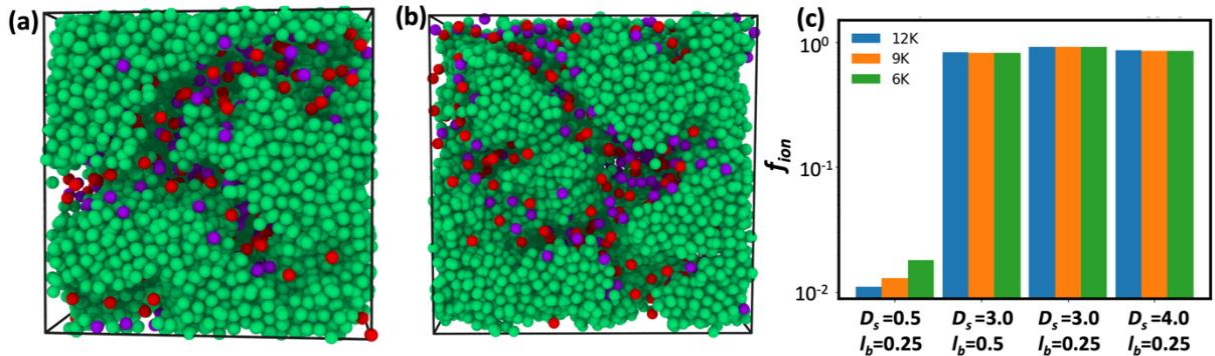

*Figure 4: Ion percolation in a polymer matrix. MD snapshot for 12k and 18 K particle-based simulations are shown in (a) and (b), respectively. In both the cases, ion pair size ratio is $D_s$=3.0 and Bjerrum length is $l_b$=0.25. The fraction of ions in the largest cluster ($f_{ion}$) for four different simulations are shown in (c).*

We further analyze the phase behavior at low ion concentrations. The phase behavior for a limiting case of $\rho = 0.025$, is shown in Figure 3. The MD snapshots in Figure 3a-e clearly suggest three phases are possible for low ion concentration viz., discrete aggregates of ions, percolated ion aggregate and macrophase separation. We recapitulate all the three phases as a function of the Bjerrum length and ion pair size ratio in Figure 3f. In the lower limit of $l_b$, the ions are microphase separated into a tortuous percolating structure within the polymer melt for higher ion size ratio. We observe discrete ion aggregates for lower ion size ratio and lower

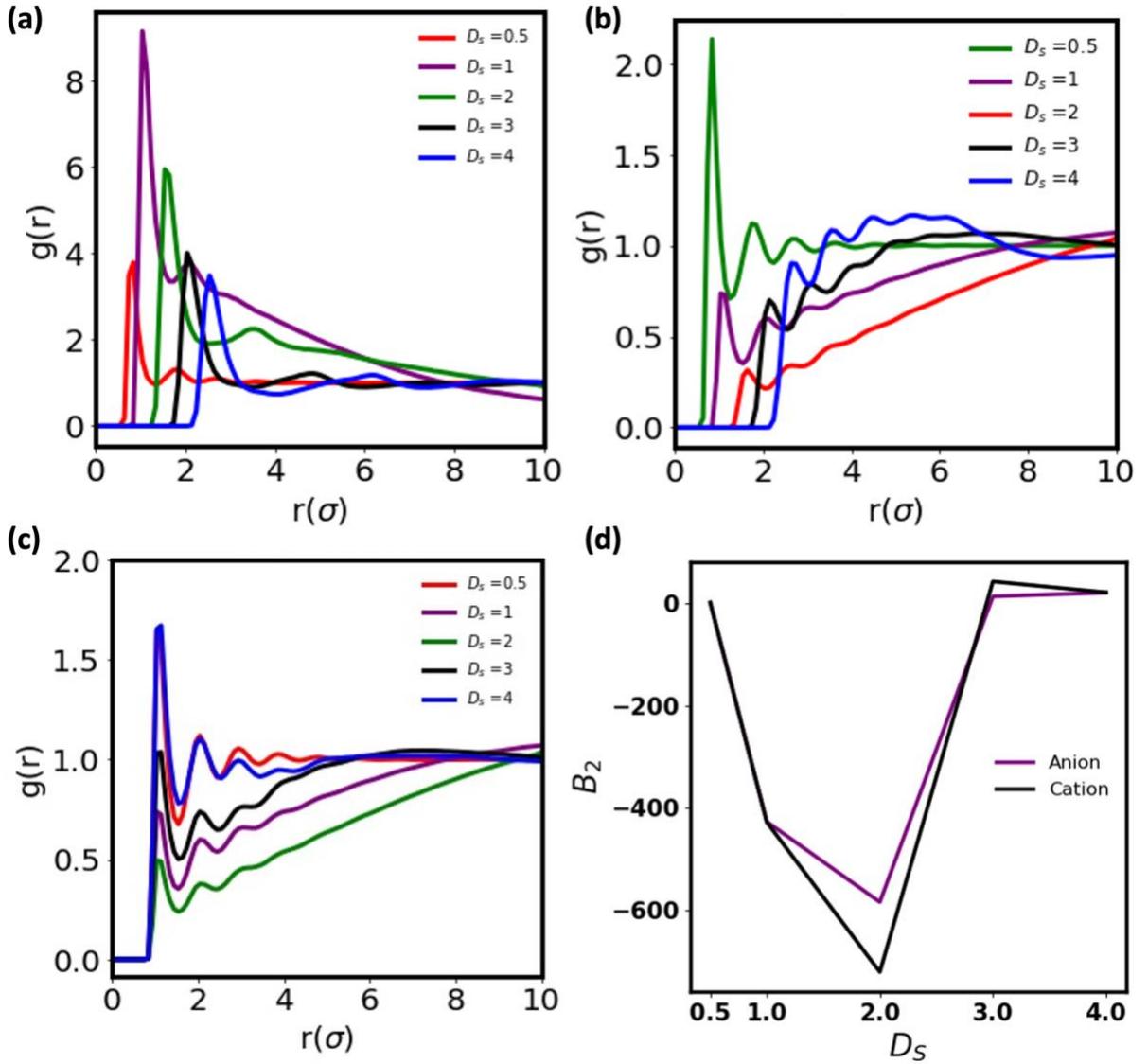

*Figure 4: Pair correlation functions of the SPE for ion concentration $\rho = 0.025$. The cation-anion, polymer-cation and polymer-anion pair correlation functions for five different ion size ratios are shown in (a), (b), and (c), respectively. The Bjerrum length $l_b=0.5$ for all the cases. The 2$^{nd}$ Virial coefficient, which is calculated from the anion-anion pair correlation function, is plotted as a function of the ion pair size ratio in (d).*

Bjerrum length. Also, the system exhibit macroscopic ion-polymer phase separation for high Bjerrum length. These two phases – discrete ion aggregates and macrophase separation are similar to the ones reported for higher concentrations (*cf.* Figure 2). To better understand the percolating structure at $\rho = 0.025$, we further conduct simulations with a varied system size. We run simulations with total number of particles varying from ~3000 to ~18000 keeping the ion concentration fixed at $\rho = 0.025$. Figure 4a and b shows the MD snapshots for 12k and 18K systems, respectively. This clearly indicates the percolation is independent of system size. The cluster analysis in Figure 4c suggest that more than 85% of ions are part of the largest aggregate that is percolating. This structure is distinctly different from the discrete aggregates of ions, which are seen in lower ion size ratio. Figure 4c indicate that the largest cluster contains only ~5% of ions of the system in case of discrete ion aggregates.

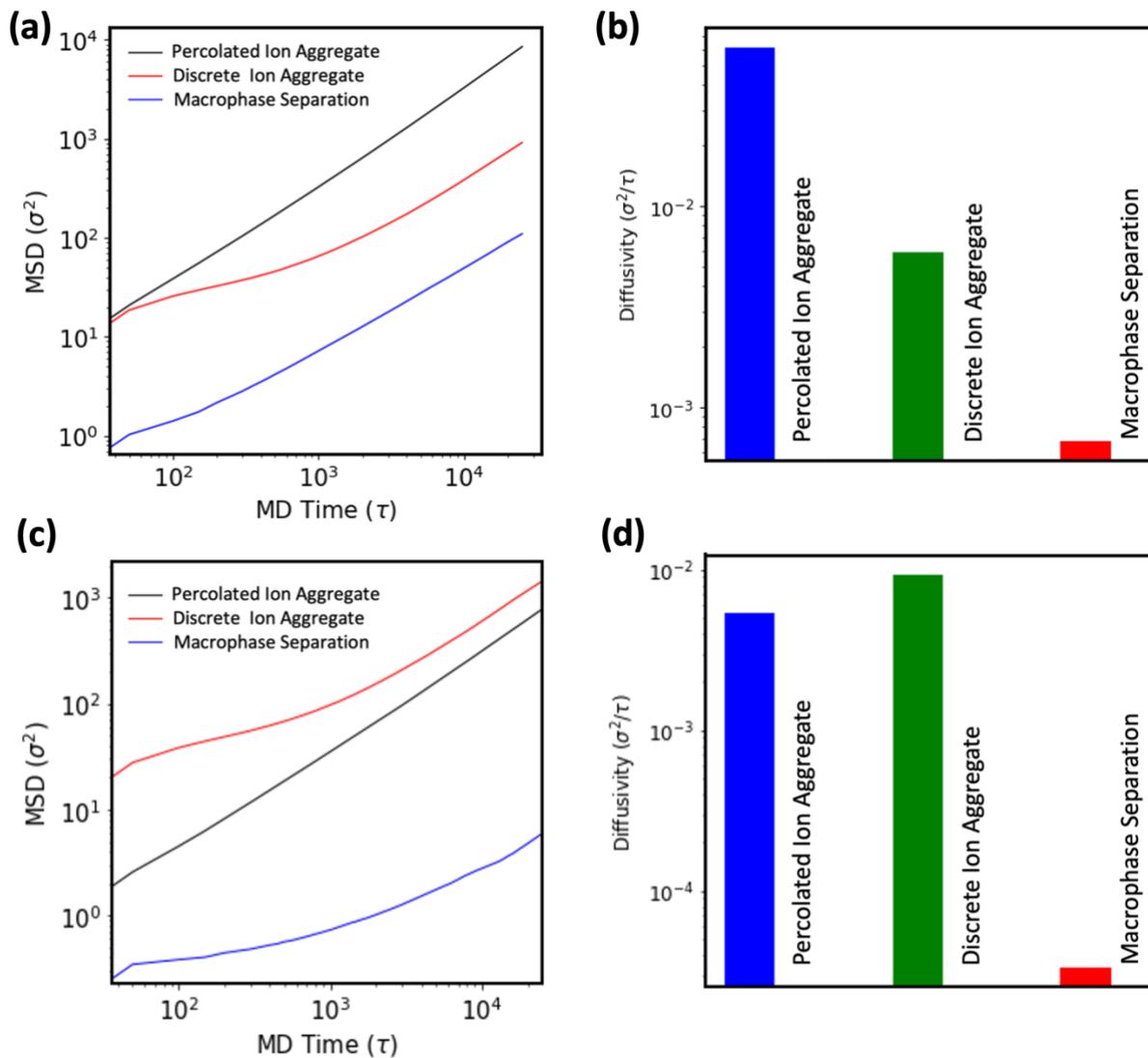

*Figure 6: (a) Mean Square Displacement (MSD) of an anion in a percolated ion aggregate phase ($l_b = 0.5$, $D_s = 3$), discrete ion aggregate phase ($l_b = 0.5$, $D_s = 1$) and macrophase separated system ($l_b = 50$, $D_s = 3$). The ion concentration for all the cases are $\rho = 0.025$. Corresponding diffusivities are shown in (b). Similarly, MSD and diffusivity of a cation in these three system are shown in (c) and (d), respectively.*

The pair correlation functions of the system are show in Figure 5. The cation-anion correlation is strongest when they are of same size as indicated by the first-peak height in Figure 5a. As the two ions are of dissimilar size, their local packing becomes weaker. For all the cases, the 1st peak position is at a interparticle distance of their mean size. This indicates there is no polymer layers in between ions, correspond to the polymer- ion microphase separation. Similarly, the ion-polymer binding and coordination are strongest when the ion pair size ratio is lowest, $D_S=0.5$. As the $D_s$ increases, the polymer-ion pair correlation decreases initially and then starts to increase when $D_S \geq 3$. This signifies a crossover from a discrete aggregation of ions to a percolating aggregation of ions in the polymer matrix. To further understand the crossover from the discrete to the percolating aggregation, we estimate the 2nd virial coefficient that define the extent of ion solvation in the polymer matrix. The 2nd virial coefficient can be defined[28] as $B_2 = -\frac{1}{2}\int_0^\infty [g_{ii}(r) - 1]4\pi r^2 dr$, wherein $g_{rr}(r)$ is the ion-ion pair correlation function. The $B_2$ is shown in Figure 5d for the citation and anion in the polymer melt for $l_b=0.5$ as a function of ion pair size ratio. The $B_2$ initially decreases as Ds increases. This correspond to good solvation of ions in the polymer melt. In the limit $D_S \geq 2.0$, the $B_2$ starts to increase. This indicate inadequate solvation of ions and lead to the long-range cluster formation.

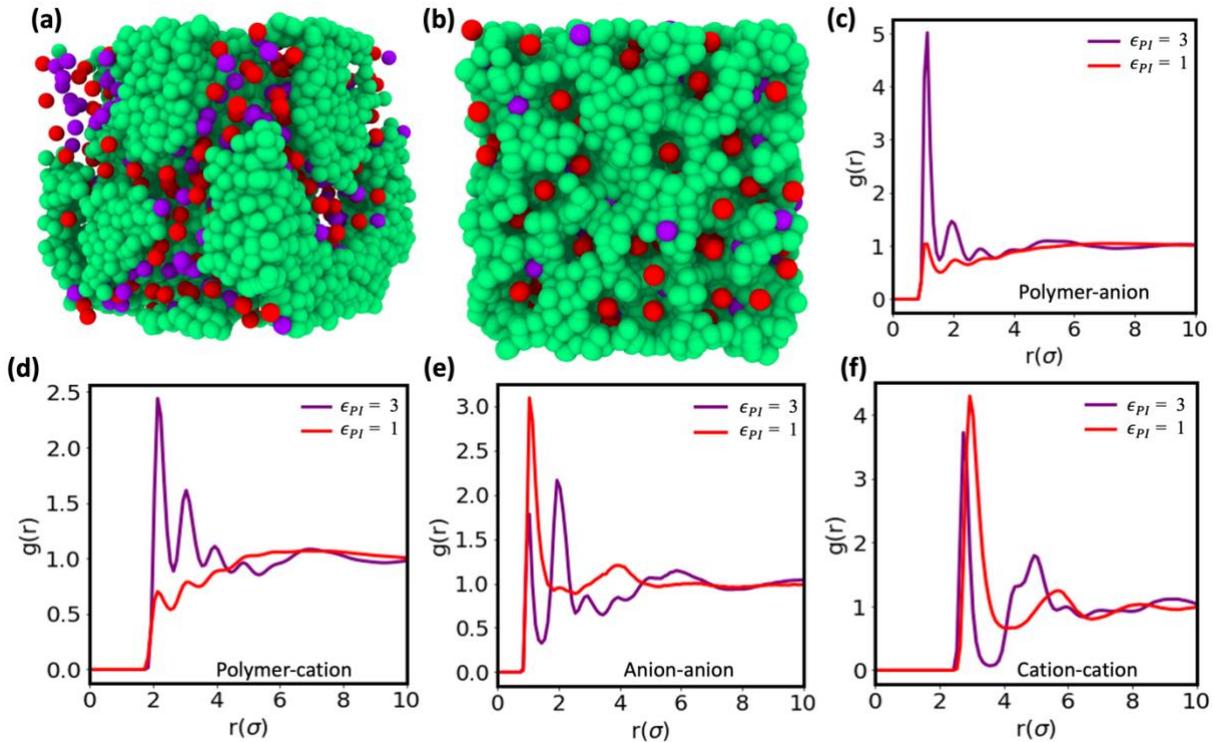

*Figure 7 : MD snapshots of the SPE for $\epsilon_{PI} = 1$ and $\epsilon_{PI} = 3$ are shown in (a) and (b), respectively. The cation and anions are shown as red and purple beads, respectively. The green beads correspond to polymers. The pair correlation functions for polymer-anion, polymer-cation, anion-anion and cation-cation are shown in (c), (d), (e), and (f), respectively for two different polymer-ion interaction strengths. All the data are for ion concentration $\rho = 0.025$.*

The means square displacement (MSD) of ions for all the three phases are shown in Figure 6 for ion concentration $\rho = 0.025$. The anion diffusion is fastest in the percolating phase, which corresponds to a greater ion pair size disparity. It is also significantly higher than the other two phases. The cation dynamics (Fig 6c,d) in percolating and discrete aggregates are comparable with a slightly higher diffusion coefficient in the discrete aggregate phase. This is in consist with experimental report on high ion conductivity for larger size cation.[29] Moreover, as the cluster size in the discrete aggregation phase increases, both the ions move faster as indicated in Figure 6. In a bigger size ion cluster, the fraction of ions that are away from the ion-polymer interface increases. This lead to the higher extent of decoupling of ion dynamics from the polymer segmental dynamics, causing faster ion movement. Overall, the discrete ion aggregation improves the cation conductivity. This is consistent with previous computational studies as well, wherein the ion-polymer interaction is modelled by dipolar interaction,[30,31] In the macrophase separated system, the ion-ion electrostatic interaction is very strong and it outperforms the polymer-solvent interaction. This corresponds to minimal ion dynamics.

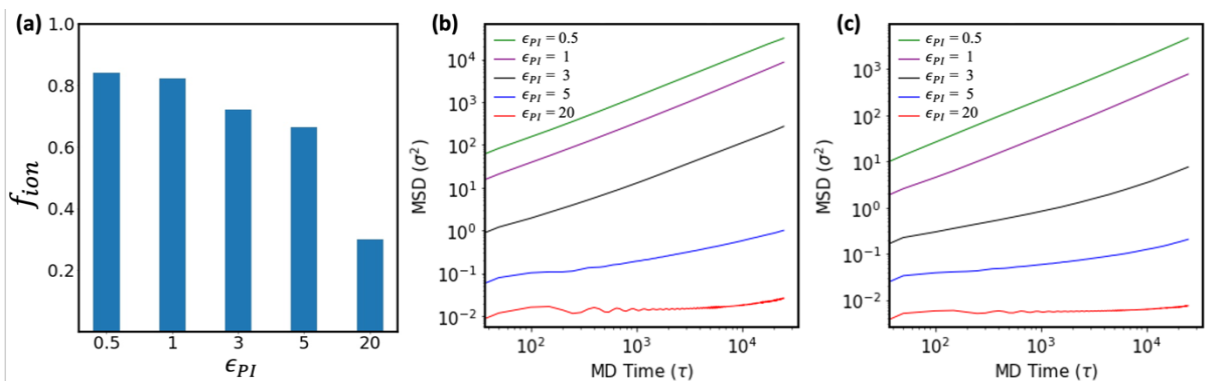

*Figure 8: The fraction of ion in the largest cluster in the SPE is plotted as a function of ion-polymer interaction in (a) for different ion-polymer interactions. The anion and cation MSD are shown in (b), and (c), respectively, for different ion-polymer interactions as a function of MD time. All the data are for ion concentration $\rho = 0.025$.*

We further study the impact of ion solvation on the phase behavior and ion dynamics. As the ion solvation in the polymer melt is determined by the polymer-ion interaction, we vary the LJ interaction strength ($\epsilon_{PI}$) between them, and analyze the microstructures of the system. The MD snapshots in Figure 7a-b suggests a fine-graining of the percolating ion domain when polymer-ion interaction increases. As the $\epsilon_{PI}$ increases further, the ion percolation supresses and smaller size ion aggregates form. The pair correlation functions clearly show an increment of ion-polymer correlations (Figure 7c-d) and an decrement of ion-ion correlation (Figure. 7e-f) as the $\epsilon_{PI}$ increases. We compute the fraction of ions in the largest clusters for different polymer-ion interactions, which is shown in Figure 8a. The fraction of ions in the largest cluster systemically decreases as the $\epsilon_{PI}$ increases. The size of the largest cluster at $\epsilon_{PI} = 0.5$ is 378.27, which is ~ 84% of the total ions in the system. This corresponds to the percolated network. However, at $\epsilon_{PI} = 20$, the largest cluster size is 135.106, which is 30 % of the total ions in the system. This represents discrete aggregates with more and more small ion clusters. This has a direct consequence on the ion dynamics. The MSD of both anion and cation are shown for the system in Figure 8b and c, respectively for $\epsilon_{PI}$ = 0.5, 1, 3,5 and 20. The dynamics of both ions become sluggish as the $\epsilon_{PI}$ increases. This analysis indicates that increased interaction between ions and polymers results in smaller ion clusters, consequently leading to reduced ion diffusion.

## IV. Conclusions

SPEs are of significant interest for various electrochemical applications, particularly in the field of energy storage devices such as lithium-ion batteries, fuel cells, and supercapacitors. Here, we study the complex correlation between ion distribution and ion dynamics in a polymer electrolyte using CGMD simulations of a phenomenological model system. We show how ion pair size ratio and the Bjerrum length of an SPE control the microstructure and ion dynamics. Particularly, we identify percolated ion aggregations for a specific range of ion pair size and the Bjerrum length of the medium for low ion concentrations. For moderately high ion concentration, the percolated aggregate becomes discrete aggregates. This phase transition strongly impact the ion dynamics. We note that a considerable body of research works on ionomers, where anion or cation is covalently bonded with polymers, demonstrate microscopic phase separation of ion domains.[32–36] The bound ions aid in controlling the ion domain formation. However, it might lead to a stronger coupling of ion motion with the polymer segmental dynamics. In the current study, we focus on SPEs, and it would be intriguing to contrast the conductivity and phase behavior of SPEs with ionomers. In summary, the current study illustrates important ion structure-dynamic correlations that have fundamental implications for the future advancement of polymer-based electrolytes.

## Acknowledgement

This work is made possible by financial support from the SERB, DST, and Govt of India through a core research grant (CRG/2022/006926) and the National Supercomputing Mission's research grant (DST/NSM/R&D_HPC_Applications/2021/40). This research uses the resources of the Center for Nanoscience Materials, Argonne National Laboratory, which is a DOE Office of Science User Facility supported under the Contract DE-AC02-06CH11357.

GKR acknowledges the Prime Minister Fellowship supported by the Ministry of Education, Gov. of India.